\documentclass[journal]{IEEEtran}
\usepackage{amsmath,amssymb,graphicx,booktabs,multirow,xcolor,cite}
\usepackage[hidelinks]{hyperref}
\usepackage{algorithm}
\usepackage{algpseudocode}

\newcommand{\ResAcceptedMain}{8}

\newcommand{\ResGbestAcc}{74.3\%}

\newcommand{\ResGbestScore}{0.784}
\newcommand{\ResFlatScore}{0.715}
\newcommand{\ResRandomScore}{0.722}
\newcommand{\ResTPEScore}{0.727}

\newcommand{\HidGeightLo}{71.6}

\newcommand{\HidGbestLo}{63.9}

\newcommand{\TinGzeroSeen}{51.2}

\newcommand{\TinGzeroPS}{0.511}

\newcommand{\TinTransferSeen}{58.1}
\newcommand{\TinTransferRay}{47.1}
\newcommand{\TinTransferPS}{0.615}

\newcommand{\TinRandSeen}{55.7}

\newcommand{\TinRandPS}{0.550}

\newcommand{\TinLastSeen}{54.5}

\newcommand{\TinLastPS}{0.539}

\begin{document}
\title{Recursive Code Evolution for Task-Oriented Semantic Communications}
\author{Yan~Gong, Zheng~Chu,~\emph{Member, IEEE}, Wei~Koong~Chai, Hari~Mohan~Pandey, Jingrui~Hou\\Miao Zhang,~\emph{Member, IEEE}, Yuanjian Li~\emph{Member, IEEE}, and Pei Xiao~\emph{Senior Member, IEEE}%
\thanks{Yan Gong, Wei Koong Chai, and Hari Mohan Pandey are with the School of Computing and Engineering, Bournemouth University, Fern Barrow, Poole, Dorset, BH12 5BB, United Kingdom (e-mail: ygong@bournemouth.ac.uk; wchai@bournemouth.ac.uk; hpandey@bournemouth.ac.uk).}%
\thanks{Zheng Chu is with the Department of Electrical and Electronic Engineering, University of Nottingham Ningbo China, Ningbo, China (e-mail: andrew.chuzheng7@gmail.com).}%
\thanks{Jingrui Hou is with the School of Information Management, Wuhan University, Wuhan, China (e-mail: houjingrui@whu.edu.cn).}%
\thanks{Miao Zhang is with the School of Information Science
and Engineering, Chongqing Jiaotong University, Chongqing 400074, China (e-mail: msczz@foxmail.com).}
\thanks{Yuanjian Li is with the Academy of Artificial Intelligence and Advanced Technology, Xi’an Jiaotong-Liverpool University, Suzhou 215123, China (e-mail: yuanjian.li@xjtlu.edu.cn)}
\thanks{Pei Xiao is with the Institute for Communication Systems (ICS), University of Surrey, Guildford, Surrey GU2 7XH, United Kingdom (e-mail: p.xiao@surrey.ac.uk).}
% \thanks{Corresponding author: Zheng Chu.}%
}
\maketitle
\begin{abstract}
Semantic communication jointly optimizes inference accuracy,
channel use, and computation, yet the underlying communication program is
typically specified before training. This paper presents ComRSI (Communications-Oriented Recursive Self-Improvement), a framework
for recursively evolving that program through large-language-model-assisted
code edits. An agent modifies the encoder, channel adapters, decoder,
loss, and training policy; a protected evaluator screens candidates under
prescribed budgets and resource constraints. Accepted code and textual
feedback inform subsequent proposals, while candidate weights are trained
from scratch. In the original ten-generation CIFAR-100 lineage, eight
accepted revisions produce an attention-pooled code achieving 74.3\%
test accuracy over an Additive White Gaussian Noise (AWGN) channel at
32 complex symbols per image, compared with 67.2\% at 256 symbols
for the initial design. 
% The program also improves fading-channel accuracy and transfers to Tiny-ImageNet after retraining. 
The program also improves fading-channel accuracy
and transfers to Tiny-ImageNet after retraining,
where a separate ComRSI search demonstrates
further recursive improvements.
A repeat study with ten searches per method compares ComRSI against non-recursive agents with and without textual history. Mean final search scores balancing
accuracy, transmission rate, and computational cost are 0.7587, 0.7376,
and 0.7089, respectively; the mean numbers of accepted revisions per search are 6.0, 5.0, and 2.1.
The advantage over the memoryless agent is supported by a two-sided rank
test, whereas the difference from the history-equipped agent remains
statistically inconclusive at this sample size. Archived proposals show
that pooling can yield a large improvement, while retained history
supports subsequent refinement. These results support recursive program
optimization within the evaluated setting and expose its remaining
rate-robustness and compute-placement trade-offs.

\end{abstract}

\section{Introduction}
\label{sec:intro}
\IEEEPARstart{S}{emantic} communication seeks to preserve the information needed by a receiver's inference task under communication and computational constraints~\cite{gunduz2023beyond,strinati2021beyondshannon,qin2021principles}. For image classification, the relevant design question is not simply how accurately an image can be reconstructed, but where semantic processing should occur and which representation should cross the channel. Encoder depth, the location of the transmitter-receiver split, representation dimension, channel adaptation, and the training objective jointly determine the resulting accuracy-rate-compute trade-off.

\begin{figure}[t]
\centering
\includegraphics[width=\columnwidth]{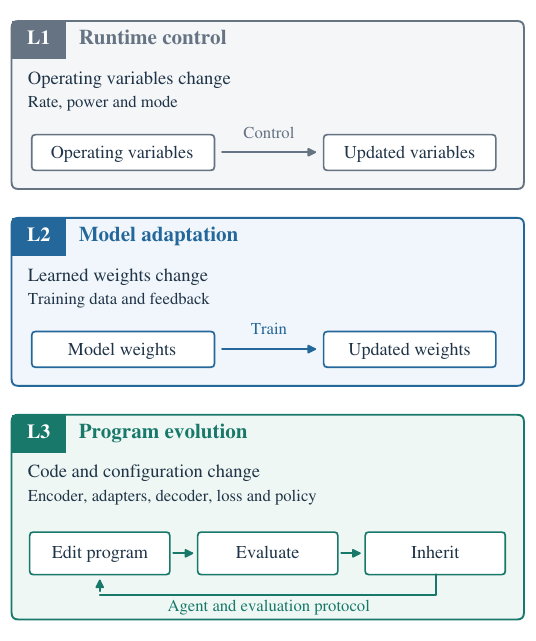}
\caption{Three complementary objects of adaptation. L1 changes operating variables, L2 changes learned weights, and L3 changes the communication program. ComRSI implements L3 through recursive code inheritance. The distinction concerns what changes, not a ranking of system intelligence.}
\label{fig:adaptation_levels}
\end{figure}

Deep Joint Source-Channel Coding (DeepJSCC) established the effectiveness of learning channel-resilient representations~\cite{bourtsoulatze2019deepjscc}. Subsequent work introduced feedback~\cite{kurka2020deepjsccf}, attention conditioned on the Signal-to-Noise Ratio (SNR)~\cite{xu2022adjscc}, adaptive rate control~\cite{yang2022adaptiverate}, and information-bottleneck formulations for edge inference~\cite{shao2022taskoriented}. These advances also illustrate the breadth of the design space: a useful improvement may require changing a representation or processing block, rather than adjusting weights or selecting a value from a fixed configuration space.

Fig.~\ref{fig:adaptation_levels} distinguishes three objects of adaptation. Runtime control changes operating variables; model adaptation updates learned weights; program evolution changes the implementation that determines what is trained and executed. These are complementary capabilities, not a universal hierarchy of communication systems. This paper studies the third capability: an agent edits a communication program, evaluates the proposed changes, and recursively builds on accepted code.

Recent agentic wireless systems automate workflow design and component orchestration~\cite{qian2024selfevolving,zhao2026agentification,tong2026wirelessagentpp,jiang2026agentictokencom}. In parallel, language-model-driven program search has produced executable solutions in mathematical and algorithmic domains~\cite{romeraparedes2024funsearch,novikov2025alphaevolve}. Applying this approach to communication design poses two challenges. First, candidates must respect the physical channel and resource accounting even when their source code changes. Second, repeated access to validation feedback can reward over-specialization to the evaluator. An increasing search score alone therefore cannot establish an improved communication system.

We introduce ComRSI (Communications-Oriented Recursive Self-Improvement), a framework that combines recursive code inheritance with a protected training and evaluation pipeline. The inherited artifact is a program specifying the semantic encoder, channel adapter, decoder, loss, and training-SNR policy. Each candidate is trained from scratch; model weights are not inherited. A fixed Large Language Model (LLM) agent proposes edits, and a fixed selector accepts a child only when its public score improves by a prescribed margin and passes a paired image-bootstrap check. Hidden test data and channel-evaluation results are withheld from the search loop and evaluated after the relevant searches are complete.

Our research question is whether this procedure can produce successive, auditable improvements in task-oriented communication and whether the selected designs retain useful accuracy-rate trade-offs beyond the public evaluation conditions. 
In ComRSI, the fixed LLM agent and the evolving communication
program are treated as components of an integrated system.
Communication is a functionality of this system, and recursively
modifying its communication program therefore constitutes
system-level self-improvement of that functionality.
Each accepted program becomes the parent of the next generation,
enabling successive improvements through inherited code
modifications. The underlying LLM agent and evaluation protocol
remain fixed; thus, the self-improvement concerns the system's
communication functionality rather than the agent's own
optimization capability.

The contributions are as follows.
\begin{itemize}
\item We introduce ComRSI, a communication-specific framework for recursive evolution of executable task-oriented semantic communication programs. Building upon general-purpose code-evolution paradigms, ComRSI integrates editable transmitter–receiver architectures and training policies with protected physical-channel constraints, resource accounting, and statistically gated evaluation.
\item We construct comparisons between recursive and non-recursive code proposals, configuration-space random search and mutation, and Bayesian optimization. The main CIFAR-100 comparisons use the same nominal proposal and training allocation. We distinguish code inheritance from textual history, extend both auxiliary searches to ten generations, and repeat the three agent variants ten times each. The analysis separates retraining variation from variation across searches.
\item We identify the communication structure produced
by the recorded search: a deeper transmitter, a pooled
semantic vector, and a small SNR-aware receiver.
The final design reaches \ResGbestAcc{} hidden-test
Additive White Gaussian Noise (AWGN) accuracy at
32 complex symbols per image, with improved
fading-channel accuracy and successful transfer
to Tiny-ImageNet. A separate ComRSI search on
Tiny-ImageNet also achieves three accepted revisions
over five generations.
\end{itemize}
The analysis also characterizes the cost of these gains: a 64-symbol ancestor is more robust at low SNR, most inference computation moves to the transmitter, and the average advantage over a non-recursive agent with textual history remains uncertain with ten searches per method.

Section~\ref{sec:related} positions the study within related work. Section~\ref{sec:method} specifies the system and search protocol. Sections~\ref{sec:setup} and~\ref{sec:results} describe the experiments and findings, and Section~\ref{sec:discussion} discusses their scope.

\section{Related Work}
\label{sec:related}
\subsection{Learned and Task-Oriented Communication}
DeepJSCC learns an encoder-decoder pair for transmission over a noisy channel~\cite{bourtsoulatze2019deepjscc}, with extensions to feedback~\cite{kurka2020deepjsccf}, channel adaptation~\cite{xu2022adjscc}, and variable-rate transmission~\cite{yang2022adaptiverate}. DeepSC~\cite{xie2021deepsc} and broader semantic communication formulations~\cite{gunduz2023beyond,strinati2021beyondshannon} emphasize meaning or task utility, while task-oriented edge inference explicitly couples representation compression and inference quality~\cite{shao2022taskoriented}. These studies primarily focus on designing and optimizing communication architectures or transmission strategies, while the systematic evolution of communication implementations through iterative code modification remains less explored.

\subsection{Agentic and Self-Evolving Wireless Systems}
Self-evolving wireless architectures seek adaptation beyond manually managed configurations~\cite{qian2024selfevolving,zhao2026agentification}. A2SSC applies agents to adaptive semantic speech communication~\cite{yang2026a2ssc}. WirelessAgent++ formulates executable agentic workflow design as program search~\cite{tong2026wirelessagentpp}, while Agentic TokenCom jointly considers workflow structures and component instances through planner, evaluator, and reflector agents~\cite{jiang2026agentictokencom}. These studies demonstrate the potential of agent-driven adaptation and workflow optimization in wireless systems. However, the recursive evolution of learned communication programs, including transmitter-receiver representations and training objectives, remains less explored, particularly under explicit physical-channel constraints and withheld evaluation conditions.

\subsection{Automated Design and Recursive Program Search}
Neural architecture search~\cite{zoph2017nas,real2019regularized} and hyperparameter optimization~\cite{bergstra2011tpe,bergstra2012random,akiba2019optuna} provide natural controls for automated communication design. Their practical results depend on the supplied search space. Language-model-based evolution~\cite{lehman2022elm}, optimizer discovery~\cite{chen2023lion}, FunSearch~\cite{romeraparedes2024funsearch}, and AlphaEvolve~\cite{novikov2025alphaevolve} instead use code as a flexible candidate representation.
A separate literature studies systems that improve their own problem-solving machinery, from theoretical self-referential programs~\cite{schmidhuber2003godel} to STOP~\cite{zelikman2023stop}, the Darwin G\"odel Machine~\cite{zhang2025dgm}, and AI research agents~\cite{lu2024aiscientist,srikanth2026rsi}. These studies explore recursive improvement through program modification, inheritance, and executable evaluation, with varying degrees of adaptation in the proposing agents themselves. Their application to learned communication systems, where program evolution must jointly account for channel robustness, representation design, and resource efficiency, remains less explored.

\section{System Model and Recursive Code Evolution}
\label{sec:method}

\subsection{Framework of ComRSI}
\begin{figure*}[t]
\centering
\includegraphics[width=\textwidth]{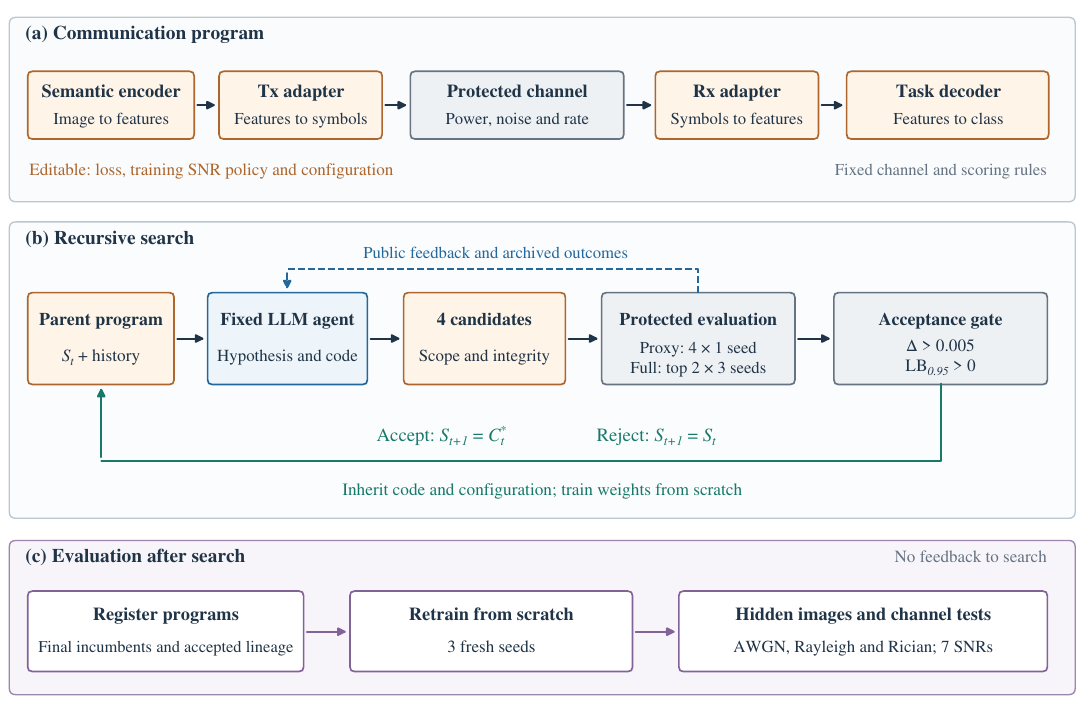}
\caption{ComRSI architecture and evaluation boundaries. (a) Editable communication modules surround a protected physical channel; the loss and training-SNR policy are also editable. (b) Four proposals are screened and the two best valid candidates are trained with three seeds; an accepted program is inherited and otherwise the incumbent is retained. Public diagnostics feed the next generation. (c) Registered programs are retrained and tested after search. Hidden results have no feedback edge to the proposal or selection loop.}
\label{fig:framework}
\end{figure*}

Fig.~\ref{fig:framework} presents the overall ComRSI framework,
which enables a task-oriented semantic communication system to
evolve through successive modifications of its executable
program. Rather than optimizing only the parameters of a
predefined neural architecture, ComRSI allows an LLM-based
agent to modify the communication system itself, including
its network structure, semantic representation, channel
adapters, training objective, and configuration. The
framework comprises three interconnected components:
an editable communication program, a recursive code
evolution process, and a protected post-search evaluation
procedure.

As shown in Fig.~\ref{fig:framework}(a), the communication
program defines how an input image is encoded into channel
symbols, transmitted through a physical channel, and
decoded into a task prediction. The semantic encoder,
transmitter and receiver adapters, task decoder, loss
function, and training-SNR policy constitute the editable
design space. In contrast, the physical channel model,
power normalization, evaluation metrics, and resource
constraints are protected from agent modifications.
This separation enables the exploration of different
communication architectures while maintaining consistent
physical-channel conditions and evaluation criteria.

The core of ComRSI is the recursive evolution process
illustrated in Fig.~\ref{fig:framework}(b). Starting from
an initial communication program $S_0$, a fixed LLM agent
examines the current program and its public evaluation
feedback, then proposes four candidate modifications.
Each candidate represents a potentially different
communication design rather than merely a new set of
trained weights. After validity screening and preliminary
training, the two highest-scoring valid candidates undergo
full-budget training with three independent seeds.
Their performance is evaluated using a composite objective
that balances task accuracy, communication rate, and
computational cost. A candidate is accepted only if its
improvement over the current program satisfies both a
minimum score margin and a paired bootstrap criterion.

Importantly, the accepted candidate becomes the parent
program for the next generation. The agent therefore
continues modifying the previously improved communication
design, using both inherited source code and accumulated
textual feedback. If no candidate satisfies the acceptance
criteria, the current program is retained while the
search history is updated. This recursive inheritance of
executable code distinguishes ComRSI from approaches that
restart each proposal from the initial architecture.
Model parameters are retrained from scratch for each
candidate and are not transferred between generations.

Finally, Fig.~\ref{fig:framework}(c) illustrates the
post-search evaluation. Selected programs are registered,
retrained with fresh random seeds, and evaluated on
withheld test images and channel conditions, including
AWGN, Rayleigh fading, and Rician fading. These results
are excluded from the proposal and acceptance loop,
providing a separate assessment of the communication
designs discovered during search. The following subsections formalize the communication model, optimization objective, recursive evolution mechanism, and evaluation safeguards. Table~\ref{tab:notations} summarizes the main mathematical notations.

\begin{table}[t]
\centering
\caption{Summary of Main Notations}
\label{tab:notations}
\setlength{\tabcolsep}{5pt}
\renewcommand{\arraystretch}{1.08}
\begin{tabular}{lp{0.67\columnwidth}}
\toprule
Notation & Description \\
\midrule
$S$, $S_0$, $S_t$ & Communication program, initial program, and incumbent at generation $t$ \\
$\boldsymbol{\theta}_{S,\xi}$ & Weights obtained by training $S$ with seed $\xi$ \\
$E$, $D$ & Semantic encoder and task decoder \\
$T_{\mathrm{tx}}$, $T_{\mathrm{rx}}$ & Transmitter and receiver channel adapters \\
$\mathbf{u}$, $\mathbf{s}$ & Unnormalized and power-normalized transmitted symbols \\
$\mathbf{r}$, $\mathbf{y}$ & Received symbols before and after equalization \\
$\mathbf{m}$, $k_a$ & Activity mask and number of active symbols \\
$h$ & Channel fading coefficient \\
$\gamma$, $\rho$ & SNR in decibels and linear scale \\
$R$, $F$ & Channel uses and forward FLOPs per image \\
$N_{\mathrm{par}}$ & Number of model parameters \\
$\mathrm{PS}_{\xi}(S)$ & PublicScore under training seed $\xi$ \\
$\overline{\mathrm{PS}}(S)$ & Seed-averaged PublicScore \\
$\mathcal{S}_{\mathrm{valid}}$ & Set of valid communication programs \\
$A$, $C_t^j$ & Fixed proposal agent and candidate $j$ at generation $t$ \\
$M_t$, $H_t$ & Public diagnostics and search history \\
$\mathcal{A}_t$ & Candidates satisfying the acceptance gate \\
$\widehat{\Delta}$ & Estimated candidate-incumbent score improvement \\
$\mathrm{LB}_{0.95}$ & One-sided 95\% bootstrap lower bound \\
\bottomrule
\end{tabular}
\end{table}

\subsection{Communication Program and Physical Channel}
Let $(\mathbf{x},c)$ denote an image and its class label. A communication program is
\begin{equation}
S=(E,T_{\mathrm{tx}},T_{\mathrm{rx}},D,L,P,\boldsymbol{\eta}),
\end{equation}
where $E$ is the semantic encoder, $T_{\mathrm{tx}}$ and $T_{\mathrm{rx}}$ are the channel adapters, $D$ is the task decoder, $L$ is the training loss, $P$ is the training-SNR policy, and $\boldsymbol{\eta}$ contains editable configuration values. A protected training procedure produces weights $\boldsymbol{\theta}_{S,\xi}=\mathcal{T}(S;\xi)$ for training seed $\xi$. Code and configuration are inherited; each evaluated program is initialized and trained afresh.

The transmitter produces $k$ complex values $\mathbf{u}$ and, optionally, an activity mask $\mathbf{m}$. At evaluation, $\mathbf{m}\in\{0,1\}^k$ and $k_a=\|\mathbf{m}\|_1$. For nonzero active energy, the protected channel forms
\begin{equation}
\mathbf{s}=\sqrt{\frac{k_a}{\|\mathbf{m}\odot\mathbf{u}\|_2^2}}\,
                (\mathbf{m}\odot\mathbf{u}),\qquad
\|\mathbf{s}\|_2^2=k_a .
\label{eq:power}
\end{equation}
The implementation includes a small numerical stabilizer in the denominator, so the energy constraint holds approximately when stabilization is active. Communication cost is $R=\mathbb{E}[k_a]$, measured in complex channel uses per image. All principal systems reported here use fixed-length codes; the optional mask is not used by the final ComRSI design.

We consider three channel models: AWGN, Rayleigh fading, and
Rician fading. Let $\gamma$ denote SNR in decibels and
$\rho=10^{\gamma/10}$ denote linear SNR. All three channels
follow the unified model
\begin{equation}
\mathbf{r}=h\mathbf{s}+\mathbf{n},\qquad
\mathbf{n}\sim\mathcal{CN}(\mathbf{0},\rho^{-1}\mathbf{I}),
\label{eq:channel}
\end{equation}
where $h$ is the channel coefficient and $\mathbf{n}$ is
complex Gaussian noise. The three channel models differ
in their definitions of $h$:

\begin{itemize}
\item \textit{AWGN:} $h=1$, representing transmission
without fading.
\item \textit{Rayleigh fading:} $h\sim\mathcal{CN}(0,1)$,
constant across all symbols of an image (block fading).
\item \textit{Rician fading:} The channel coefficient is
\begin{equation}
h=\sqrt{\frac{K_{\mathrm{ric}}}{K_{\mathrm{ric}}+1}}e^{j\phi}
 +\sqrt{\frac{1}{K_{\mathrm{ric}}+1}}g ,
\end{equation}
where $K_{\mathrm{ric}}=3$, $\phi$ is uniform on
$[0,2\pi)$, and $g\sim\mathcal{CN}(0,1)$.
\end{itemize}

For Rayleigh and Rician fading, the protected receiver
applies perfect-CSI linear minimum mean-square error
(LMMSE) equalization under a unit-symbol-power assumption:
\begin{equation}
\mathbf{y}=\mathbf{m}\odot
 \frac{h^*\mathbf{r}}{|h|^2+\rho^{-1}},\qquad
\gamma_{\mathrm{eff}}=10\log_{10}(|h|^2\rho).
\end{equation}
For AWGN, $\mathbf{y}=\mathbf{m}\odot\mathbf{r}$ and
$\gamma_{\mathrm{eff}}=\gamma$.
The receiver uses the equalized symbols and channel metadata
to predict the class. Channel-state acquisition, signaling
overhead, and digital modulation are outside the
evaluated model.

The channel module passes the nominal SNR and
$\gamma_{\mathrm{eff}}$ as metadata. The final $G_{10}$
receiver uses $\gamma_{\mathrm{eff}}$ to construct its
two SNR features, while the transmitter does not
condition on SNR.

\subsection{Public Objective and Feasible Programs}
We define PublicScore (PS) as a composite evaluation
metric combining classification accuracy, communication
rate, and computational cost. For validation data
$\mathcal{D}_{\mathrm{val}}$ and public AWGN grid
$\Gamma=\{-3,0,3,6,9\}$\,dB, it is calculated as
\begin{align}
\mathrm{PS}_{\xi}(S)
&=\frac{1}{|\Gamma|}\sum_{\gamma\in\Gamma}
 \mathrm{Acc}_{\mathrm{val}}(\boldsymbol{\theta}_{S,\xi},\gamma)
 \nonumber\\
&\quad-\alpha\ln\frac{R_{\xi}(S)}{R_0}
      -\beta\ln\frac{F_{\xi}(S)}{F_0},
\label{eq:ps}
\end{align}
where $F$ is the measured number of Forward Floating-Point Operations (FLOPs) per image, $(R_0,F_0)$ are frozen references for the initial design $S_0$, and $(\alpha,\beta)=(0.02,0.01)$. Accuracy is a fraction, not a percentage. The score is a scalar design utility; it is not an accuracy measure. A twofold rate reduction contributes $0.02\ln2\approx0.0139$, equivalent to 1.39 accuracy percentage points, whereas doubling FLOPs costs approximately 0.69 points. These coefficients were fixed before the search and assign twice the logarithmic penalty weight to rate as to forward computation.

Search seeks a high seed-averaged score, defined as $\overline{\mathrm{PS}}(S)=\mathbb{E}_{\xi}[\mathrm{PS}_{\xi}(S)]$ and estimated using three training seeds, within the valid program set:
\begin{equation}
\max_{S\in\mathcal{S}_{\mathrm{valid}}}\ \overline{\mathrm{PS}}(S)
\quad\text{s.t.}\quad
N_{\mathrm{par}}(S)\leq1.5N_{\mathrm{par},0},\
F(S)\leq2F_0 .
\end{equation}
Validity includes interface, code-scope, numerical, and integrity checks. The implementation searches this set heuristically; it supplies no global-optimality guarantee. Because training steps are fixed but candidate architectures differ, equal training allocations do not imply equal training FLOPs or wall-clock time.

% \begin{figure*}[t]
% \centering
% \includegraphics[width=\textwidth]{figures/fig2_comrsi_framework.pdf}
% \caption{ComRSI architecture and evaluation boundaries. (a) Editable communication modules surround a protected physical channel; the loss and training-SNR policy are also editable. (b) Four proposals are screened and the two best valid candidates are trained with three seeds; an accepted program is inherited and otherwise the incumbent is retained. Public diagnostics feed the next generation. (c) Registered programs are retrained and tested after search. Hidden results have no feedback edge to the proposal or selection loop.}
% \label{fig:framework}
% \end{figure*}

\subsection{Proposal, Screening, and Inheritance}
At generation $t$, the agent receives parent program $S_t$, public diagnostics $M_t$, and archived history $H_t$. With fixed proposal agent $A$,
\begin{equation}
\{C_t^1,\ldots,C_t^4\}=A(S_t,M_t,H_t).
\end{equation}
Diagnostics include per-SNR accuracy, rate, FLOPs, class-level errors, confidence, and training-loss summaries. The agent examines public performance diagnostics to
identify potential weaknesses in the current communication
program. It then formulates four distinct modification
hypotheses and implements them as candidate code revisions.
Each proposal includes a rationale linking the intended
modification to the observed performance. The resulting
candidates are independently evaluated, allowing the
selection process to determine which modifications
provide measurable improvements.
The main runs use the archived model identifier \texttt{claude-opus-5-5}. 

Candidates passing scope and interface checks receive one proxy training run. The top two valid candidates by proxy score are trained from scratch at the full budget using seeds $\Xi=\{0,1,2\}$. They are compared against the same incumbent, using cached evaluations where available. Let $\mathcal{A}_t$ contain the candidates satisfying the acceptance gate below. Then
\begin{equation}
S_{t+1}=
\begin{cases}
\displaystyle\arg\max_{C\in\mathcal{A}_t}\overline{\mathrm{PS}}(C),
 &\mathcal{A}_t\neq\varnothing,\\
S_t,&\mathcal{A}_t=\varnothing .
\end{cases}
\label{eq:inherit}
\end{equation}
The selected program is the sole parent of the next generation. Candidate branches, diffs, hypotheses, and accepted ancestry are archived. A rejected generation changes the history but not the inherited program. The complete recursive evolution procedure is summarized in Algorithm~\ref{alg:comrsi}.

\begin{algorithm}[t]
\caption{ComRSI Evolution Framework}
\label{alg:comrsi}
\begin{algorithmic}[1]
\Require Initial program $S_0$, fixed agent $A$,
public evaluator, generation budget $G$
\Ensure Final communication program $S_G$
\State Initialize $S \gets S_0$, history $H \gets \varnothing$,
and public diagnostics $M$ from $S_0$
\For{$t=0,\ldots,G-1$}
    \State Generate four candidates using $A(S,M,H)$
    \State Screen candidates for validity and resource constraints
    \State Proxy-train valid candidates and rank by PublicScore
    \State Select up to two top-ranked valid candidates
    \State Train selected candidates with seeds $\{0,1,2\}$
    \State Evaluate candidate-incumbent score differences
    \State Apply paired-bootstrap acceptance gate
    \If{at least one candidate is accepted}
        \State $S \gets$ highest-scoring accepted candidate
    \EndIf
    \State Update public diagnostics $M$ and history $H$
\EndFor
\State \Return $S$
\end{algorithmic}
\end{algorithm}

\subsection{Paired Acceptance Gate and Its Interpretation}
For validation image $i$, let $q_i(S)$ be correctness averaged over the five public SNRs and three full-budget training seeds. Let $b(S)$ denote the seed-averaged rate and compute terms in~\eqref{eq:ps}. For a candidate $C$ and incumbent $S_t$,
\begin{equation}
\widehat{\Delta}(C,S_t)=
 \frac{1}{n_{\mathrm{val}}}\sum_i[q_i(C)-q_i(S_t)]
 +b(C)-b(S_t).
\end{equation}
The selector draws 2,000 paired bootstrap resamples of validation-image indices~\cite{efron1979bootstrap}, preserving each candidate-incumbent pairing. The fifth percentile of the resampled score differences is the one-sided lower bound $\mathrm{LB}_{0.95}$. Acceptance requires
\begin{equation}
\widehat{\Delta}>0.005,\qquad
\mathrm{LB}_{0.95}>0,
\label{eq:accept}
\end{equation}
and valid full-budget evaluations for all three seeds.

This gate reduces acceptance of small, image-sampling-sensitive
changes. Hidden evaluation and fresh training seeds provide
additional checks of selected programs, while ten independent
searches per agent variant assess search-level variability.

\subsection{Protected Evaluation and Information Flow}
The data manifests, channel, trainer, evaluator, selector, and budget constants are protected by SHA-256 checks before a generation and after agent proposals. The agent operates in a separate workspace with tool-level access restrictions and a permitted candidate checker. Static checks restrict imports and reject prohibited I/O and RNG reseeding. 

Integrity checks additionally replace the received code with noise and test permutation equivariance over the inference batch. For $C$ classes, the noise-only accuracy threshold is $\max(5/C,1/C+0.05)$; normalized transmitter and receiver permutation errors must be below $10^{-2}$. A cheating fixture that stores encoder features for later receiver access is detected by the repository test. 

Search feedback contains public results only.
Before hidden evaluation, a registration file records
the evaluation targets, including all accepted ComRSI
generations and comparator incumbents.
The additional repetitions and auxiliary experiments
use the same benchmark holdout, and hidden scores
are never provided to the proposal agents.

\section{Experimental Protocol}
\label{sec:setup}
\subsection{Data and Initial Communication System}
CIFAR-100~\cite{krizhevsky2009cifar} is the primary benchmark: 45,000 official training images are used for training, 5,000 for public validation, and the 10,000 official test images for hidden evaluation. Tiny-ImageNet~\cite{le2015tinyimagenet} provides 200 classes and $64\times64$ images; the corresponding split uses 90,000 training images, 10,000 public validation images, and the 10,000 official validation images as the hidden set. Fixed stratified manifests define the splits. 

The initial system $S_0$ has a ResNet-style~\cite{he2016resnet} transmitter with a stem and three residual stages of widths of 64, 128, and 256. A $1\times1$ projection produces an $8\times8\times8$ real-valued latent, corresponding to $R_0=256$ complex symbols. The receiver lifts the latent to 256 channels, applies two residual blocks, and performs global pooling and classification. For $64\times64$ inputs, a stride-two stem preserves the latent dimensions. On CIFAR-100, $S_0$ has 3.63 million parameters and 0.694 Giga Floating-Point Operations (GFLOPs) per image. It uses cross-entropy loss and a uniform training-SNR policy on $[-3,9]$\,dB.

\subsection{Compared Methods}
\label{sec:baselines}
We compare ComRSI with human-designed references, non-recursive
agent variants, configuration-space search methods, and
restricted agent variants. The main agent-based experiments
use \texttt{claude-opus-5-5} as the proposal model, except
for the earlier-agent experiment, which uses \texttt{claude-sonnet-5-5}. All agent variants share the same public
evaluator and acceptance gate. The methods differ in
how candidates are generated, how previous evaluation
results are used, and whether accepted program code
is inherited across generations. Importantly, retaining
the best accepted program for evaluation does not
necessarily imply inheriting its code for subsequent
proposals.

\noindent\textit{Initial system and hand tuning.}
The initial system $S_0$ is the human-designed
communication architecture described above and serves
as the common starting point for automated searches.
Hand-tuned $S_0$ selects the best configuration from
ten configurations evaluated over two tuning rounds,
while retaining the original architecture. These
references establish the performance of the initial
communication design before automated optimization.

\noindent\textit{ComRSI (ours).}
At each generation, the Opus agent directly modifies
the current accepted communication program using
its source code, previous proposal history, and
public evaluation feedback. Candidate modifications
may change network structures, semantic representations,
channel adapters, training objectives, or configuration
values. Accepted changes become part of the inherited
program for subsequent generations. Thus, ComRSI
accumulates improvements through both executable
code inheritance and textual history, while model
weights are retrained from scratch.

\noindent\textit{Non-Recursive Agent with History (NR-H).}
This baseline uses the same Opus model and evaluation
procedure as ComRSI, but starts every generation
from the original program $S_0$. The agent retains
textual descriptions of previous proposals and their
public evaluation outcomes, allowing earlier ideas
to inform subsequent edits. However, accepted source
code is not directly inherited. The external selector
still retains the best accepted program for evaluation
and comparison. This baseline examines the contribution
of direct code inheritance when textual history
remains available. NR-H and ComRSI share the editable interfaces and nominal evaluation budgets; their distinguishing feature is the inheritance of accepted program source code.

\noindent\textit{Non-Recursive Agent without History (NR-0).}
This variant of NR-H also starts each generation from $S_0$,
but receives no accumulated proposal or evaluation
history, and access to previous candidate workspaces
is denied. Each generation therefore produces
candidates without inherited code or textual
feedback from earlier generations. The external
selector nevertheless tracks the best accepted
program across generations, allowing fair comparison
of final incumbents. This baseline evaluates the
search process without either form of inter-generation
inheritance.

\noindent\textit{Random mutation.}
This method randomly modifies configuration values
of the current accepted program within a predefined
search space. Accepted modifications become the
starting configurations for subsequent generations,
whereas rejected proposals leave the incumbent
unchanged. It therefore preserves recursive
inheritance of configuration changes but replaces
agent-directed source-code modification with random
configuration proposals. This comparison examines
whether recursive inheritance alone is sufficient
to achieve improvements comparable to ComRSI.

\noindent\textit{Random search.}
This method independently samples candidate
configurations from the predefined search space,
including architectural and training hyperparameters.
Each new configuration is generated without using
previous evaluation outcomes to guide sampling.
The external selector retains the best accepted
configuration, but earlier candidates do not become
parents of subsequent proposals. Unlike ComRSI,
the method explores a fixed configuration space
rather than generating unrestricted program edits.

\noindent\textit{Optuna TPE.}
This method uses the tree-structured Parzen
estimator~\cite{bergstra2011tpe,akiba2019optuna}
to guide configuration search. Unlike random
search, TPE uses the configurations and scores
of previous evaluations to model promising
regions of the predefined search space and
select subsequent candidates. It therefore
incorporates historical performance information
into the proposal process, but does not recursively
inherit modified program source code or introduce
new operations outside the specified configuration
space.

\noindent\textit{Earlier-agent run.}
This variant follows the same recursive
code-evolution procedure as ComRSI but replaces
\texttt{claude-opus-5-5} with the archived Sonnet
proposal model (\texttt{claude-sonnet-5-5}). The editable interfaces, public
evaluator, acceptance mechanism, and inheritance
rules remain unchanged. This experiment examines
the sensitivity of recursive communication-program
evolution to the choice of LLM proposal model.
The original search was subsequently continued
to ten generations, as described below.

\noindent\textit{Restricted-file agent.}
This variant uses the Opus agent and recursive
code inheritance but restricts editable files
to the training loss, training-SNR policy, and
configuration files. Unlike unrestricted ComRSI,
the agent cannot directly rewrite the encoder,
channel-adapter, or decoder implementation files.
However, editable configuration values may still
change network width, latent dimension, or other
architecture-related settings. This comparison
examines the effect of restricting the scope
of program modification rather than eliminating
architectural adaptation entirely.

The predefined configuration space for random
mutation, random search, and Optuna TPE includes
network width, depth, latent channels,
SNR-FiLM~\cite{perez2018film},
squeeze-and-excitation~\cite{hu2018se},
label smoothing, optimizer settings, and
training-SNR bounds. These methods optimize within
a predefined configuration space, whereas
agent-driven code evolution allows modifications
to the communication architecture, representation,
and training procedure.

\subsection{Training and Search Budgets}
The protected trainer uses Stochastic Gradient Descent (SGD) with Nesterov momentum 0.9, batch size 256, 3\% linear warm-up followed by cosine decay, gradient-norm clipping at 5, bfloat16 autocast, and random-crop/flip augmentation. Learning rate and weight decay remain editable within protected ranges. Each CIFAR-100 proxy uses 2,500 optimizer steps and each full training uses 10,000 steps, approximately 57 passes through the training set. Tiny-ImageNet doubles both step budgets. All candidate models are trained and publicly evaluated
under AWGN channels. Rayleigh and Rician fading channels
are reserved for post-search evaluation to assess
robustness under unseen channel models.

All eight CIFAR-100 search procedures are
evaluated over ten generations, with four proposals per generation,
one proxy seed, and three full-budget seeds for
each of two promoted candidates. This corresponds
to a common nominal candidate-training allocation
of $10[4(2500)+2(3)(10000)]=700{,}000$ optimizer
steps per search, excluding initial-system evaluation,
agent checker runs, and post-search retraining.

All methods follow the same nominal training-step
allocation. All experiments
were conducted on a single NVIDIA RTX PRO 6000 GPU
with 96\,GB of memory.

\subsection{Withheld Evaluation Conditions and Retraining}

After the search process, the initial and hand-tuned
systems, final comparator incumbents, and all accepted
ComRSI generations are retrained from scratch using
three fresh random seeds (100-102). The resulting
models are evaluated on withheld test images under
AWGN, Rayleigh, and Rician fading channels at
$\Gamma_{\mathrm{hid}}=\{-6,-3,0,3,6,9,12\}$\,dB.
The public and hidden channel random seeds are
disjoint.

Evaluation includes PublicScore, classification
accuracy, communication rate, model parameter counts,
and computational complexity. Search PublicScore
is obtained using training seeds 0-2, while
fresh-seed PublicScore is computed after retraining
with seeds 100-102. Both use the public validation
set and the objective in~\eqref{eq:ps}. Hidden-test
accuracy is reported separately. AWGN-5 averages
classification accuracy over
$\{-3,0,3,6,9\}$\,dB, while Rayleigh-7 and
Rician-7 average accuracy over all seven SNR values
in $\Gamma_{\mathrm{hid}}$. Reported fresh-seed
results are averaged over three retrainings.

For Tiny-ImageNet, the communication program
discovered on CIFAR-100 is transferred by adapting
the input dimensions and class count, followed by
training from scratch on Tiny-ImageNet. Separately, ComRSI is run for five generations on Tiny-ImageNet, starting from the initial communication program $S_0$.

\section{Results and Analysis}
\label{sec:results}

\subsection{Search Trajectory and Retraining Stability}

\begin{figure*}[t]
\centering
\includegraphics[width=\textwidth]{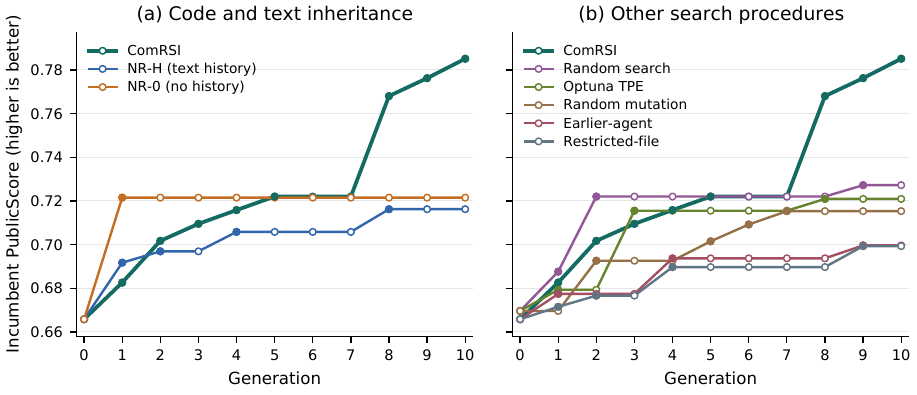}
\caption{Evolution of incumbent PublicScore over ten generations.
(a) Comparison of ComRSI with non-recursive agents with
(NR-H) and without (NR-0) textual history.
(b) Comparison with other search procedures.
Filled and hollow markers indicate accepted revisions
and retained incumbents, respectively.
Higher PublicScore indicates better performance.}
\label{fig:generations}
\end{figure*}

Fig.~\ref{fig:generations} illustrates the evolution
of incumbent PublicScore over ten generations.
ComRSI accepts \ResAcceptedMain{} revisions,
including five consecutive improvements at G1-G5,
followed by further improvements at G8-G10.
Its search score increases from 0.666 to 0.785,
while retraining the final program with three
fresh seeds yields a PublicScore of \ResGbestScore{},
demonstrating consistent performance across
training seeds.

The accepted revisions progressively improve the
communication program. G1 and G3 adjust regularization
and the training-SNR range, while G2 and G5 deepen
the semantic encoder. G4 introduces label smoothing.
A major structural change occurs at G8, where the
spatial latent representation is replaced by a
64-symbol pooled semantic code and a new receiver.
G9 further reduces the code length to 32 symbols,
and G10 introduces learned attention pooling.
These successive modifications illustrate how
ComRSI evolves the communication architecture
and training strategy through recursive code
inheritance.

\begin{figure*}[t]
\centering
\includegraphics[width=\textwidth]{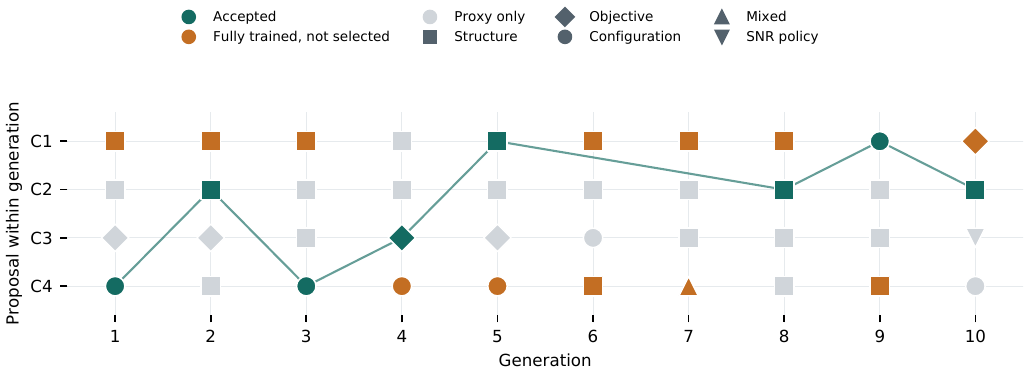}
\caption{Candidate-level evolution and code inheritance
in ComRSI over ten generations. C1-C4 denote the
four proposals per generation. Colors indicate
accepted, fully trained but rejected, and proxy-only
candidates, while shapes represent modification types.
The connecting line traces the accepted code lineage.}
\label{fig:lineage}
\end{figure*}

Fig.~\ref{fig:lineage} further illustrates the
candidate-level evolution of ComRSI. Each generation
contains four proposed modifications, of which
selected candidates undergo full training and
evaluation. The accepted candidates form a continuous
code-inheritance lineage across generations, while
rejected candidates do not replace the incumbent.
This illustrates how ComRSI progressively evolves
the communication program through selective
inheritance of successful code modifications.

\subsection{The Communication Structure Produced by Search}
The recursive search progressively modifies the communication
program, with a major architectural transition occurring
at G8 and further refinement at G9-G10. This subsection
examines the final communication architecture discovered
by ComRSI. Its encoder produces
$\mathbf{V}=[\mathbf{v}_1,\ldots,\mathbf{v}_{N}]\in\mathbb{R}^{d\times N}$
with $d=256$ and $N=64$ spatial locations. Three learned attention heads and a global-average branch form pooled features:
\begin{align}
a_{\ell j}
 &=\frac{\exp(\mathbf{w}_{\ell}^{\mathsf T}\mathbf{v}_j+b_\ell)}
 {\sum_{r=1}^{N}\exp(\mathbf{w}_{\ell}^{\mathsf T}\mathbf{v}_r+b_\ell)},\\
\mathbf{p}_0&=\frac{1}{N}\sum_j\mathbf{v}_j,\qquad
\mathbf{p}_\ell=\sum_j a_{\ell j}\mathbf{v}_j,\quad \ell=1,2,3,\\
\mathbf{z}&=\mathbf{W}_{p}
 [\mathbf{p}_0^{\mathsf T},\ldots,\mathbf{p}_3^{\mathsf T}]^{\mathsf T}
 +\mathbf{b}_{p}\in\mathbb{R}^{2k}.
\label{eq:pool}
\end{align}
The $2k$ real outputs are paired into $k=32$ complex symbols and normalized by~\eqref{eq:power}. Attention logits are zero-initialized, so the learned pools begin as uniform averages.

At the receiver, the real and imaginary parts of the channel outputs are concatenated with two SNR features:
\begin{equation}
\mathbf{v}_{\mathrm{rx}}=
 [\operatorname{realvec}(\mathbf{y})^{\mathsf T},
   \gamma_{\mathrm{eff}}/10,\ 10^{-\gamma_{\mathrm{eff}}/20}]^{\mathsf T}.
\end{equation}
A linear layer, batch normalization, and ReLU produce a 512-dimensional hidden vector, followed by a linear classifier. The resulting design has 5.29 million parameters and 1.14 GFLOPs, within the protected caps. Its structural consequence is a shift from transmitting a spatial feature map to transmitting a compact task representation after substantial local processing. The study discovers this design within the recorded search; it does not establish attention pooling as a new primitive.

\subsection{Inheritance and Configuration-Space Comparisons}

Table~\ref{tab:methods} compares ComRSI with non-recursive
agents and configuration-space search methods.
ComRSI achieves a fresh-seed PublicScore of
\ResGbestScore{}, compared with \ResFlatScore{}
for NR-H and 0.722 for NR-0. NR-0 accepts a pooled
128-symbol program at G1 and retains it through G10,
whereas NR-H accepts four revisions and ends at
384 symbols.

\begin{table*}[t]
\centering
\caption{Comparison of communication-program search methods
on CIFAR-100 over ten generations.
Search PS and Fresh-seed PS denote public evaluation scores
obtained during search and after retraining, respectively.
Fresh-seed PS is reported as mean $\pm$ standard deviation
over three retrainings. $R$ denotes complex channel uses
per image, Params (M) denotes millions of parameters,
and GFLOPs denotes forward computation per image.}
\label{tab:methods}
\setlength{\tabcolsep}{7pt}
\begin{tabular}{lrrrrrr}
\toprule
Method & Accepted revisions & Search PS & Fresh-seed PS & $R$ & Params (M) & GFLOPs \\
\midrule
$S_0$ (human design) & - & - & 0.669$\pm$0.004 & 256 & 3.63 & 0.69 \\
Hand-tuned $S_0$ & - & - & 0.686$\pm$0.001 & 256 & 3.63 & 0.69 \\
\midrule
ComRSI (ours) & 8 & 0.785 & 0.784$\pm$0.002 & 32 & 5.29 & 1.14 \\
NR-H (text history) & 4 & 0.716 & 0.715$\pm$0.002 & 384 & 5.12 & 1.00 \\
NR-0 (no history) & 1 & 0.722 & 0.722$\pm$0.001 & 128 & 4.94 & 0.55 \\
Random mutation & 4 & 0.715 & 0.710$\pm$0.002 & 384 & 4.01 & 1.00 \\
Random search & 3 & 0.727 & 0.722$\pm$0.006 & 256 & 4.00 & 1.00 \\
Optuna TPE & 3 & 0.721 & 0.727$\pm$0.003 & 384 & 4.06 & 1.00 \\
Earlier-agent run & 3 & 0.700 & 0.704$\pm$0.002 & 256 & 5.23 & 1.02 \\
Restricted-file agent & 4 & 0.699 & 0.699$\pm$0.001 & 320 & 5.17 & 1.18 \\
\bottomrule
\end{tabular}
\end{table*}

Among the configuration-space methods, random mutation
achieves a fresh-seed PublicScore of 0.710, compared
with \ResRandomScore{} for random search and
\ResTPEScore{} for Optuna TPE. These methods improve
the initial communication system, while ComRSI
explores a broader program-editing space that
enables structural modifications such as the
pooled-code architecture discovered at G8.

Relative to the hand-tuned reference, the final
ComRSI program improves fresh-seed PublicScore
by 0.098 while reducing channel uses from
256 to 32. Compared with Optuna TPE,
ComRSI achieves a higher fresh-seed PublicScore
(0.784 versus 0.727) with fewer channel uses
(32 versus 384).

The earlier-agent variant achieves a fresh-seed
PublicScore of 0.704, compared with 0.784 for
the main ComRSI run. Restricting the set of editable source files further yields a PublicScore of
0.699. These results suggest that both the
proposal agent and the scope of executable
program modifications influence search outcomes.

\subsection{Hidden-Image and Fading-Channel Performance}

Table~\ref{tab:hidden} presents the hidden-test
classification accuracy under AWGN, Rayleigh,
and Rician fading channels. 
The final ComRSI program achieves 74.3\% AWGN-5
accuracy using only 32 complex symbols per image,
compared with 67.2\% and 256 symbols for the
initial system. The G8 ancestor achieves stronger
low-SNR and fading-channel performance, reaching
67.1\% under Rayleigh fading compared with
64.0\% for G10. In contrast, G10 halves the
communication rate while maintaining comparable
AWGN-5 accuracy. These results highlight the
trade-off between communication efficiency and
channel robustness within the discovered
program lineage.

\begin{table*}[ht]
\centering
\caption{CIFAR-100 hidden-test classification accuracy
(\%) averaged over three fresh training seeds.
AWGN-5 averages $\{-3,0,3,6,9\}$\,dB,
while Rayleigh-7 and Rician-7 average
$\{-6,-3,0,3,6,9,12\}$\,dB.
$R$ denotes complex channel uses per image.}
\label{tab:hidden}
\setlength{\tabcolsep}{8pt}
\begin{tabular}{lrrrrrr}
\toprule
System & $R$ & AWGN $-6$\,dB & AWGN-5 & AWGN $12$\,dB & Rayleigh-7 & Rician-7 \\
\midrule
$S_0$ (human design) & 256 & 48.5 & 67.2 & 70.9 & 39.9 & 43.6 \\
Hand-tuned $S_0$ & 256 & 51.0 & 68.5 & 72.2 & 42.0 & 45.9 \\
Random search & 256 & 70.3 & 72.7 & 73.1 & 50.6 & 55.2 \\
Optuna TPE & 384 & 70.3 & 72.8 & 73.2 & 50.6 & 55.2 \\
Random mutation & 384 & 68.6 & 72.1 & 72.7 & 48.1 & 52.5 \\
NR-H (text history) & 384 & 69.9 & 72.0 & 72.4 & 53.9 & 58.8 \\
NR-0 (no history) & 128 & 61.5 & 69.8 & 71.5 & 49.8 & 54.4 \\
Earlier-agent run & 256 & 59.0 & 70.4 & 72.8 & 47.6 & 52.0 \\
Restricted-file agent & 320 & 56.4 & 70.6 & 73.4 & 41.9 & 45.9 \\
\midrule
ComRSI $G_5$ & 256 & 69.3 & 72.2 & 72.8 & 50.8 & 55.5 \\
ComRSI $G_8$ (ancestor) & 64 & 71.6 & 74.2 & 74.6 & 67.1 & 70.7 \\
ComRSI $G_{10}$ (final) & 32 & 63.9 & 74.3 & 75.0 & 64.0 & 68.4 \\
\bottomrule
\end{tabular}
\end{table*}

\begin{figure*}[!ht]
\centering
\includegraphics[width=\textwidth]{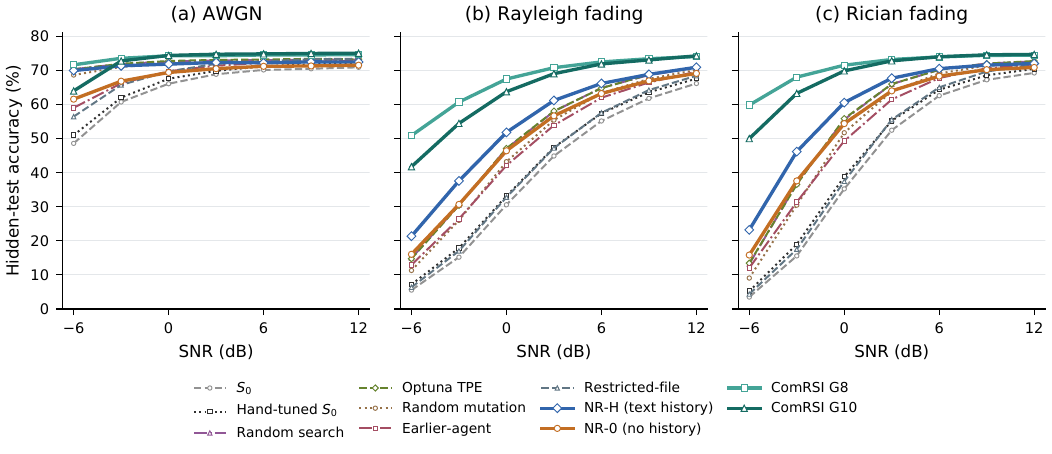}
\caption{CIFAR-100 hidden-test classification accuracy
versus SNR under AWGN, Rayleigh, and Rician fading
channels, averaged over three independent retrainings.}
\label{fig:hidden}
\end{figure*}

Fig.~\ref{fig:hidden} further illustrates
classification accuracy as a function of SNR
under the three channel models.
The SNR-dependent results show that the
discovered programs maintain strong accuracy
across a range of channel conditions.
G8 demonstrates particularly strong performance
at low SNRs and under fading, while G10
achieves competitive overall accuracy
with a more compact 32-symbol representation.
The improvements reflect the combined effects
of transmitter-side feature processing,
semantic pooling, and SNR-aware decoding.

\begin{figure}[ht]
\centering
\includegraphics[width=\columnwidth]{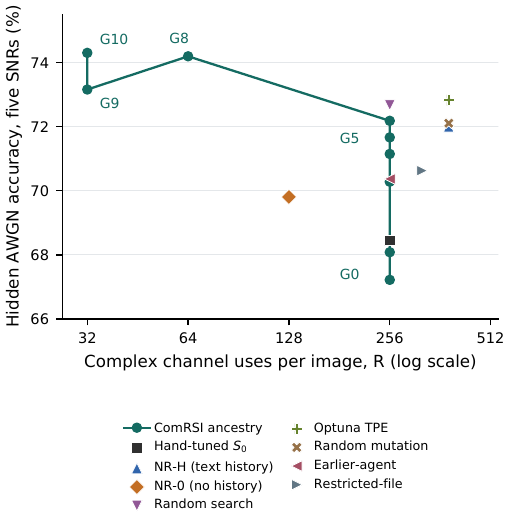}
\caption{Trade-off between hidden-test AWGN-5 accuracy
and the number of transmitted complex symbols
per image. The horizontal axis uses a logarithmic
scale.}
\label{fig:pareto}
\end{figure}

\subsection{Rate, Robustness, and Compute Placement}

Fig.~\ref{fig:pareto} illustrates the trade-off
between communication rate and classification
accuracy. The early accepted revisions improve
accuracy without reducing the transmission rate,
while G8 introduces a compact 64-symbol
representation. G9 further reduces the rate to
32 symbols, and G10 improves accuracy through
learned attention pooling.

Reducing the transmission rate from 64 to
32 symbols contributes approximately 0.0139
to PublicScore through the rate penalty term.
Although G8 achieves higher accuracy at
$-6$\,dB AWGN (\HidGeightLo{}\% versus
\HidGbestLo{}\% for G10), G10 offers a
more compact representation while maintaining
comparable average AWGN accuracy.

The evolved design shifts computational complexity
toward the transmitter, enabling a lightweight
receiver while reducing communication overhead.
For the initial $S_0$ program, the transmitter and receiver require 1.227 and 2.406 million parameters, with 0.390 and 0.304 GFLOPs per image, respectively. In the final G10 program, these become 5.203 and 0.086 million parameters, with 1.145 GFLOPs at the transmitter and 0.17 MFLOPs at the receiver. FLOPs are counted for convolutional and matrix-multiplication operations, with each multiply-add counted as two operations. This evolution concentrates semantic processing at the transmitter and produces a compact receiver alongside the 32-symbol code.

\subsection{Program Transfer and Search on Tiny-ImageNet}

We further evaluate the transferability of
communication programs discovered on CIFAR-100
and the effectiveness of direct ComRSI search
on Tiny-ImageNet. All models are trained from
scratch on Tiny-ImageNet.

\begin{table}[ht]
\centering
\caption{Communication-program transfer and search
results on Tiny-ImageNet. C100 and TIN indicate
programs discovered on CIFAR-100 and Tiny-ImageNet,
respectively. PS denotes fresh-seed PublicScore,
while accuracy (\%) is evaluated on the hidden
test set. Results are averaged over three training
runs with different random seeds. $R$ denotes
complex symbols per image.}
\label{tab:tin}
\setlength{\tabcolsep}{3pt}
\resizebox{\columnwidth}{!}{%
\begin{tabular}{llcccccc}
\toprule
System & Source & $R$ & PS & AWGN-5 & $-6$\,dB & Ray.-7 & Ric.-7 \\
\midrule
$S_0$ & - & 256 & 0.511 & 51.2 & 33.2 & 31.4 & 34.3 \\
Hand-tuned $S_0$ & C100 & 256 & 0.519 & 51.9 & 35.1 & 32.2 & 35.2 \\
Random search best & C100 & 256 & 0.550 & 55.7 & 51.4 & 38.5 & 41.9 \\
NR-H best & C100 & 384 & 0.542 & 56.1 & 52.4 & 38.2 & 41.8 \\
ComRSI $G_{10}$ & C100 & 32 & 0.615 & 58.1 & 43.7 & 47.1 & 51.1 \\
\midrule
ComRSI-TIN $G_{1}$ & TIN & 256 & 0.517 & 52.1 & 36.4 & 33.3 & 36.4 \\
ComRSI-TIN $G_{3}$ & TIN & 256 & 0.523 & 52.9 & 37.5 & 34.0 & 37.3 \\
ComRSI-TIN $G_{4}$ & TIN & 256 & 0.539 & 54.5 & 40.7 & 35.4 & 38.7 \\
\bottomrule
\end{tabular}}
\end{table}

Table~\ref{tab:tin} shows that the transferred
ComRSI G10 program achieves a fresh-seed
PublicScore of \TinTransferPS{} and hidden
AWGN-5 accuracy of \TinTransferSeen{}\%,
outperforming the transferred random-search
program (\TinRandPS{} and \TinRandSeen{}\%).
It also achieves \TinTransferRay{}\% under
Rayleigh fading while requiring only 32
complex symbols per image.

A separate five-generation ComRSI search
on Tiny-ImageNet accepts three revisions,
improving fresh-seed PublicScore from
\TinGzeroPS{} to \TinLastPS{} and hidden
AWGN-5 accuracy from \TinGzeroSeen{}\%
to \TinLastSeen{}\%. The transferred CIFAR-100
G10 program achieves higher performance than
the programs discovered within this shorter
Tiny-ImageNet search, demonstrating the
transferability of the evolved communication
architecture across the two datasets.

\subsection{Repeatability Across Ten Searches per Method}
\label{sec:repeat}

\subsubsection{Experimental Setup}

We conduct ten independent ten-generation searches
for each of ComRSI, NR-H, and NR-0 on CIFAR-100,
using the same proposal model, candidate allocation,
training budgets, and acceptance criteria.
Each final communication program is retrained
from scratch using three random seeds (100-102)
and evaluated under the same hidden-test protocol.
A complete ten-generation search constitutes
one independent experimental run.

\begin{table*}[ht]
\centering
\caption{Results over ten independent searches per
method. Values are reported as mean $\pm$
sample standard deviation across searches.
Fresh-seed metrics average three retrainings
within each search. Two-sided Mann-Whitney tests comparing ComRSI
with NR-0 and NR-H yield $p=0.0010$ and
$p=0.1212$, respectively, for final search
PublicScore.}
\label{tab:repeat}
\setlength{\tabcolsep}{12pt}
\begin{tabular}{lccc}
\toprule
Metric & ComRSI & NR-H (text history) & NR-0 (no history) \\
\midrule
Final search PS & 0.7587$\pm$0.0242 & 0.7376$\pm$0.0275 & 0.7089$\pm$0.0219 \\
Fresh-seed PS & 0.7567$\pm$0.0242 & 0.7362$\pm$0.0274 & 0.7082$\pm$0.0210 \\
Hidden AWGN-5 accuracy (\%) & 74.0$\pm$1.3 & 72.5$\pm$1.2 & 69.3$\pm$1.1 \\
Hidden Rayleigh-7 accuracy (\%) & 60.6$\pm$6.9 & 53.4$\pm$9.9 & 49.9$\pm$5.3 \\
Hidden Rician-7 accuracy (\%) & 65.0$\pm$6.8 & 57.6$\pm$10.1 & 54.3$\pm$5.5 \\
\midrule
Mean accepted revisions & 6.0 & 5.0 & 2.1 \\
Mean gain beyond the largest step & 0.0553 & 0.0447 & 0.0133 \\
Final rate below 256 symbols & 7/10 & 5/10 & 6/10 \\
Pooled-vector adoption (archive audit) & 7/10 & 4/10 & 5/10 \\
\bottomrule
\end{tabular}
\end{table*}

\subsubsection{Overall Performance and Continued Improvement}

Table~\ref{tab:repeat} shows that ComRSI achieves
the highest mean final search PublicScore (0.7587),
compared with 0.7376 for NR-H and 0.7089 for NR-0.
This ranking is maintained after fresh-seed
retraining and across all three hidden-channel
evaluations. ComRSI improves hidden AWGN-5
accuracy by 1.5 and 4.7 percentage points
over NR-H and NR-0, respectively.

% The Mann-Whitney tests yield $p=0.1212$
% for ComRSI versus NR-H and $p=0.0010$
% for ComRSI versus NR-0. The improvement
% over NR-0 is statistically significant,
% while the observed advantage over NR-H
% does not reach the 5\% significance level.

\begin{figure*}[!t]
\centering
\includegraphics[width=\textwidth]{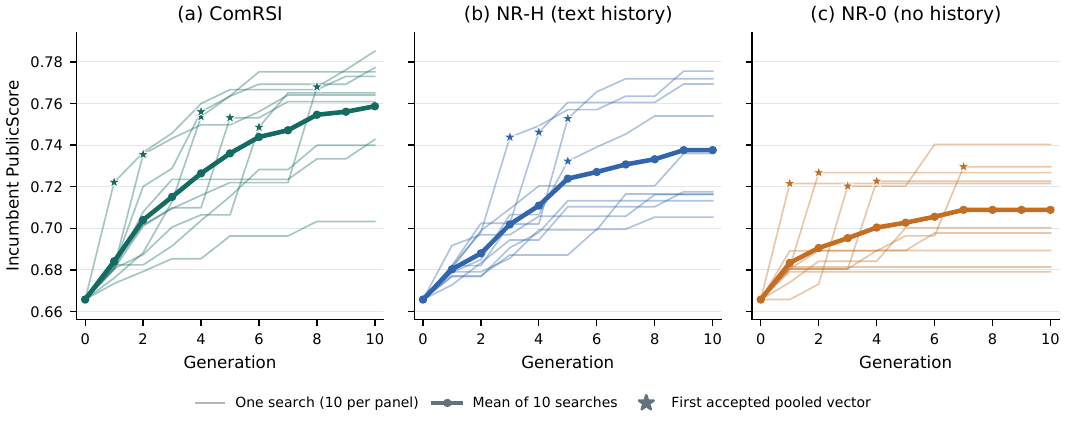}
\caption{PublicScore evolution across ten
independent searches for ComRSI, NR-H, and NR-0.
Thin lines represent individual searches,
while thick lines show the mean trajectories.
Stars indicate the first accepted pooled-vector
design in each search.}
\label{fig:repeat}
\end{figure*}

Fig.~\ref{fig:repeat} illustrates the evolution
of incumbent PublicScore across the ten
independent searches for each method.
ComRSI accepts an average of 6.0 revisions,
compared with 5.0 for NR-H and 2.1 for NR-0.
To quantify improvements beyond a single
major revision, we define
\begin{equation}
G_{\mathrm{rest}}=
(s_{10}-s_0)-
\max_{1\leq t\leq10}(s_t-s_{t-1}),
\end{equation}
where $s_t$ denotes the incumbent PublicScore
at generation $t$.
ComRSI achieves the highest mean
$G_{\mathrm{rest}}$ of 0.0553, compared with
0.0447 for NR-H and 0.0133 for NR-0.
These results indicate that ComRSI accumulates
performance improvements through successive
accepted revisions.

\subsubsection{Pooling Transitions and Subsequent Refinement}

The repeated searches also reveal differences
in architectural discovery and refinement.
Pooled-vector representations emerge in
7/10 ComRSI searches, 4/10 NR-H searches,
and 5/10 NR-0 searches.
At the first accepted pooling revision,
the mean PublicScore improvements are
0.0443, 0.0415, and 0.0447, respectively,
indicating that pooling produces substantial
gains across all three methods.

The main distinction appears in subsequent
refinement. All ComRSI searches that discover
pooling achieve further improvements (7/7),
as do all corresponding NR-H searches (4/4),
compared with only 1/5 for NR-0.
The mean additional PublicScore gains after
pooling are 0.0234, 0.0239, and 0.0040,
respectively.

These results highlight the importance of
retaining previous discoveries for continued
optimization. ComRSI directly inherits
accepted program code, while NR-H can
reuse previous discoveries through textual
history. In contrast, NR-0 generates new
proposals without access to earlier
search outcomes.

\section{Discussion and Limitations}
\label{sec:discussion}

\subsection{Discussion}

The experimental results demonstrate the potential
of recursive code evolution for task-oriented semantic
communication. ComRSI progressively improves executable
communication programs by inheriting accepted modifications,
enabling cumulative refinement of model architectures,
training strategies, and communication configurations.
The discovered 32-symbol design achieves higher
classification accuracy and substantially lower
communication overhead than the initial system,
while maintaining strong performance under fading
channels.

The repeated experiments further demonstrate the
value of retaining previous discoveries during
communication-program optimization. ComRSI achieves
the highest mean PublicScore and hidden-test accuracy
across ten searches per method. Both ComRSI and NR-H
continue improving after discovering pooled semantic
representations, highlighting the importance of
accumulated code and textual knowledge. The advantage
of ComRSI over NR-0 is statistically significant,
while its mean advantage over NR-H has not reached
statistical significance.

The discovered 32-symbol and 64-symbol designs also
illustrate the trade-off between communication
efficiency and channel robustness. This highlights
the flexibility of the proposed framework, where
different communication objectives and constraints
can guide recursive program evolution toward
different operating requirements.

\subsection{Limitations and Future Directions}

The current evaluation focuses on image classification
under simulated wireless channels. Future extensions
may consider diverse communication tasks, realistic
channel conditions, and deployment-aware optimization
incorporating channel estimation, pilot overhead,
digital modulation, finite-blocklength coding, and
adaptive rates under latency and energy constraints.

Future work may incorporate structured reasoning memory
to reuse successful and unsuccessful communication-program
editing experiences~\cite{ouyang2025reasoningbank}.
Joint evolution of the communication program and its
proposal agent offers another direction~\cite{xu2026evo2}.

\section{Conclusion}
This paper presented ComRSI, a framework for
recursively evolving executable programs for
task-oriented semantic communication.
By inheriting accepted code modifications,
ComRSI enables progressive optimization of
communication architectures, training strategies,
and transmission configurations under protected
evaluation constraints.

Experiments on CIFAR-100 demonstrate substantial
improvements in classification accuracy and
communication efficiency, with the discovered
32-symbol program outperforming the 256-symbol
initial design. The evolved program also exhibits robustness under fading channels and transferability to Tiny-ImageNet, where a separate search demonstrates further improvements. Repeated experiments across ten
independent searches per method further show that
ComRSI achieves the highest mean performance among
the evaluated search approaches.

These findings highlight recursive code evolution
as a promising approach to automated semantic
communication design. Future work will extend
the framework toward multi-task communication,
adaptive channel-aware optimization, and
deployment-oriented communication systems.

\bibliographystyle{IEEEtran}
\bibliography{references}
\end{document}